\documentclass{ws-procs10x7}
\usepackage{balance}

\newcolumntype{d}[1]{D{.}{.}{#1}}

\def\Journal#1#2#3#4{{\it #1} {\bf #2}, #3 (#4)}

\makeindex
\begin{document}

\title{Simulation of jet quenching in heavy ion collisions}

\author{I. P. LOKHTIN$^*$ \and A. M. SNIGIREV}

\address{Skobeltsyn Institute of Nuclear Physics, Moscow State University, 
Moscow, 119992, Russia 
\\$^*$E-mail: Igor.Lokhtin@cern.ch}


\twocolumn[\maketitle\abstract{
The method to simulate the rescattering and energy loss of hard partons in 
heavy ion collisions has been developed. The model is a fast 
Monte-Carlo tool introduced to modify a standard PYTHIA jet event. The full 
heavy ion event is obtained as a superposition of a soft hydro-type state and 
hard multi-jets. The model is applied to analysis of the jet quenching pattern 
at RHIC and LHC.} 
]

\section{Introduction}

One of the important tools for studying the properties of quark-gluon plasma 
(QGP) in ultrarelativistic heavy ion collisions is the analysis of a QCD jet 
production. The medium-induced energy loss of energetic partons, ``jet 
quenching'', should be very different in the cold nuclear matter and QGP, 
resulting in many observable phenomena~\cite{baier_rev}. Recent RHIC data on 
high-p$_T$ particle production at $\sqrt{s}=200~A$ GeV are in agreement with the 
jet quenching hypothesis (see, e.g.,~\cite{Wang:2004} and references therein). At 
LHC, a new regime of heavy ion physics will be reached at 
$\sqrt{s_{\rm NN}}=5.5 A$ TeV where hard and semi-hard particle production can 
stand out against the underlying soft events. The initial gluon densities in 
Pb+Pb reactions at LHC are expected to be much higher than those at RHIC, 
implying a stronger partonic energy loss, observable in new channels.

In the most of available Monte-Carlo heavy ion event generators the 
medium-induced partonic rescattering and energy loss are either ignored or 
implemented insufficiently. Thus, in order to analyze RHIC data on high-p$_T$ 
hadron production and test the sensitivity of LHC observables to the QGP 
formation, the development of adequate and fast Monte-Carlo tool to simulate the
jet quenching is necessary.  

\section{Physics model and simulation procedure} 

The detailed description of physics model can be found in our recent 
paper~\cite{lokhtin-model}. The approach bases on an accumulating energy loss, 
the gluon radiation being associated with each parton scattering in the 
expanding medium and includes the interference effect using the modified 
radiation spectrum $dE/dl$ as a function of decreasing temperature $T$. The 
basic kinetic integral equation for the energy loss $\Delta E$ as a function of 
initial energy $E$ and path length $L$ has the form 
\begin{equation} 
\label{elos_kin}
\Delta E (L,E) = \int\limits_0^L\frac{dl}{\lambda}\frac{dP}{dl}
\frac{dE}{dl} ,~ 
\frac{dP}{dl} = \frac{{\rm e}^{-l/\lambda}}{\lambda}
\end{equation} 
where $l$ is the current transverse coordinate of a parton, $dP/dl$ is the 
scattering probability density, $dE/dl$ is the energy loss per unit length, 
$\lambda$ is in-medium mean free path. The collisional loss in high-momentum 
transfer limit and radiative loss in BDMS approximation~\cite{baier} (with 
``dead-cone'' generalization of the radiation spectrum for heavy 
quarks~\cite{dc}) are using. We consider realistic nuclear 
geometry~\cite{lokhtin00} and treat the medium in nuclear overlapping zone as a 
boost-invariant longitudinally expanding quark-gluon fluid. The model 
parameters are the initial conditions for the QGP formation for central Au+Au 
(Pb+Pb) collisions at RHIC (LHC): the proper formation time $\tau_0$ and 
the temperature $T_0$. For non-central collisions we suggest the 
proportionality of the initial energy density $\varepsilon _0$ to the ratio of 
nuclear overlap function and transverse area of nuclear 
overlapping~\cite{lokhtin00}. The simple Gaussian parameterization of gluon 
angular distribution over the emission angle $\theta$ with the typical angle of 
the coherent radiation $\theta _0 \sim 5^0$~\cite{lokhtin98}. 

The model has been constructed as the Monte-Carlo event generator PYQUEN 
(PYthia QUENched) and is available via Internet~\cite{pyquen}. The routine is 
implemented as a modification of the standard PYTHIA$\_$6.4 jet 
event~\cite{pythia}. The event-by-event simulation procedure includes the 
generation of the initial parton spectra with PYTHIA and production vertexes 
at given impact parameter, rescattering-by-rescattering simulation of the 
parton path length in a dense zone, radiative and collisional energy loss per 
rescattering, final hadronization with the Lund string model for hard partons 
and in-medium emitted gluons.

The full heavy ion event is simulated as a superposition of soft hydro-type 
state and hard multi-jets. The simple approximation~\cite{lokhtin-model} of 
hadronic liquid at ``freeze-out'' stage has been used to treat soft part of the 
event. Then the hard part of the event includes PYQUEN multi-jets generated 
according to the binomial distribution. The mean number of jets produced in AA 
events at a given impact parameter is a product of the number of binary NN 
sub-collisions and the integral cross section of hard process in $pp$ 
collisions with the minimal transverse momentum transfer $p_T^{\rm min}$. The 
extended in such a way model has been also constructed as the fast Monte-Carlo 
event generator~\cite{hydjet}. Note that ideologically similar approximation has 
been developed in~\cite{hirano}. 

\section{Jet quenching at RHIC}

\begin{figure}[b]
\centerline{\psfig{file=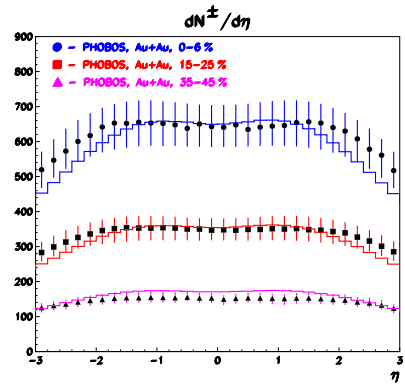,width=2.2in}}
\caption{The pseudorapidity distribution of charged hadrons in Au+Au 
collisions for three centrality sets. The points are PHOBOS data, 
histograms are the model calculations.}
\label{fig1}
\end{figure}

\begin{figure}[b]
\centerline{\psfig{file=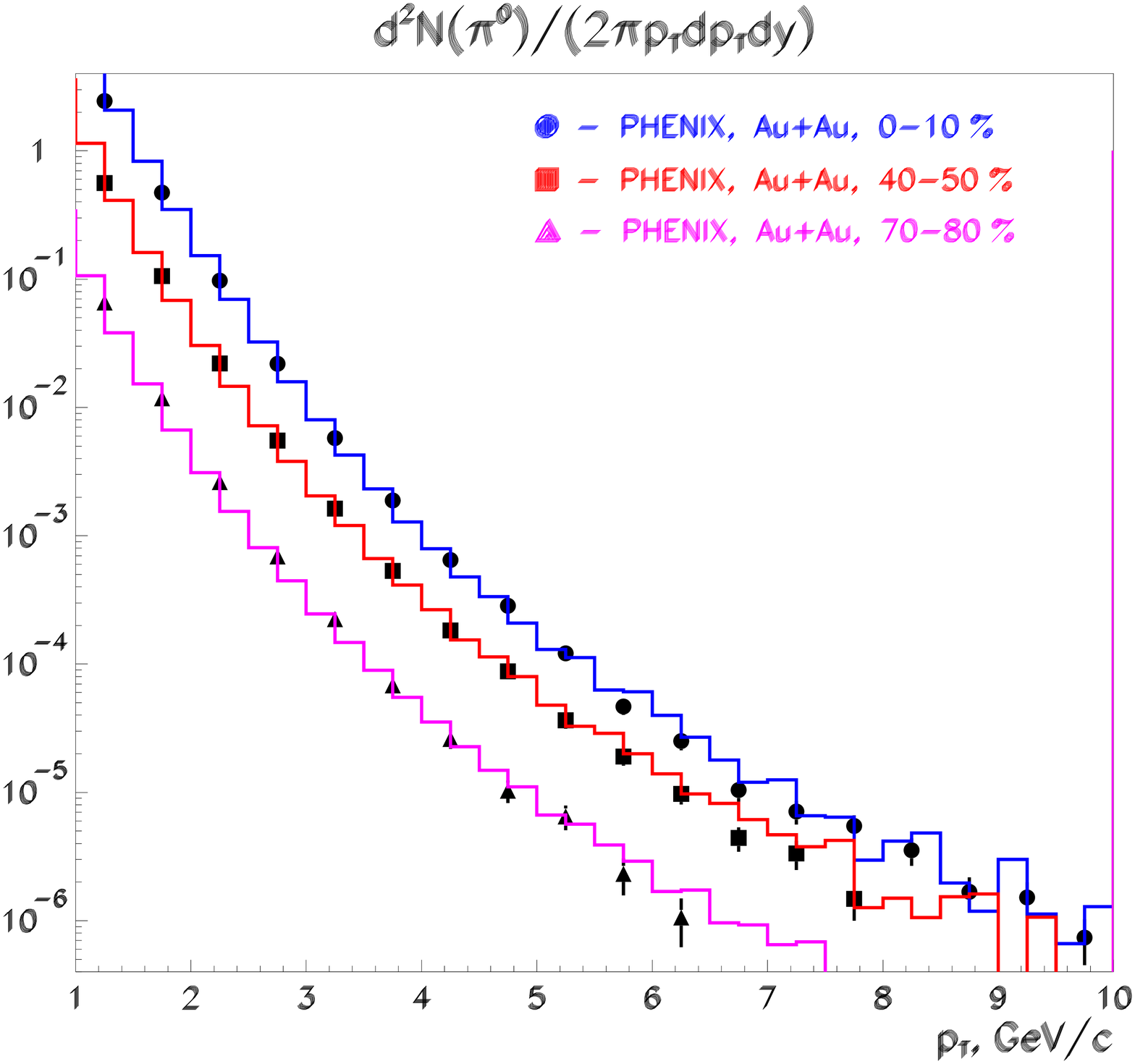,width=2.2in}}
\caption{The transverse momentum distribution of neutral pions in Au+Au 
collisions for three centrality sets. The points are PHENIX data, 
histograms are the model calculations.}
\label{fig2}
\end{figure}

\begin{figure}[b]
\centerline{\psfig{file=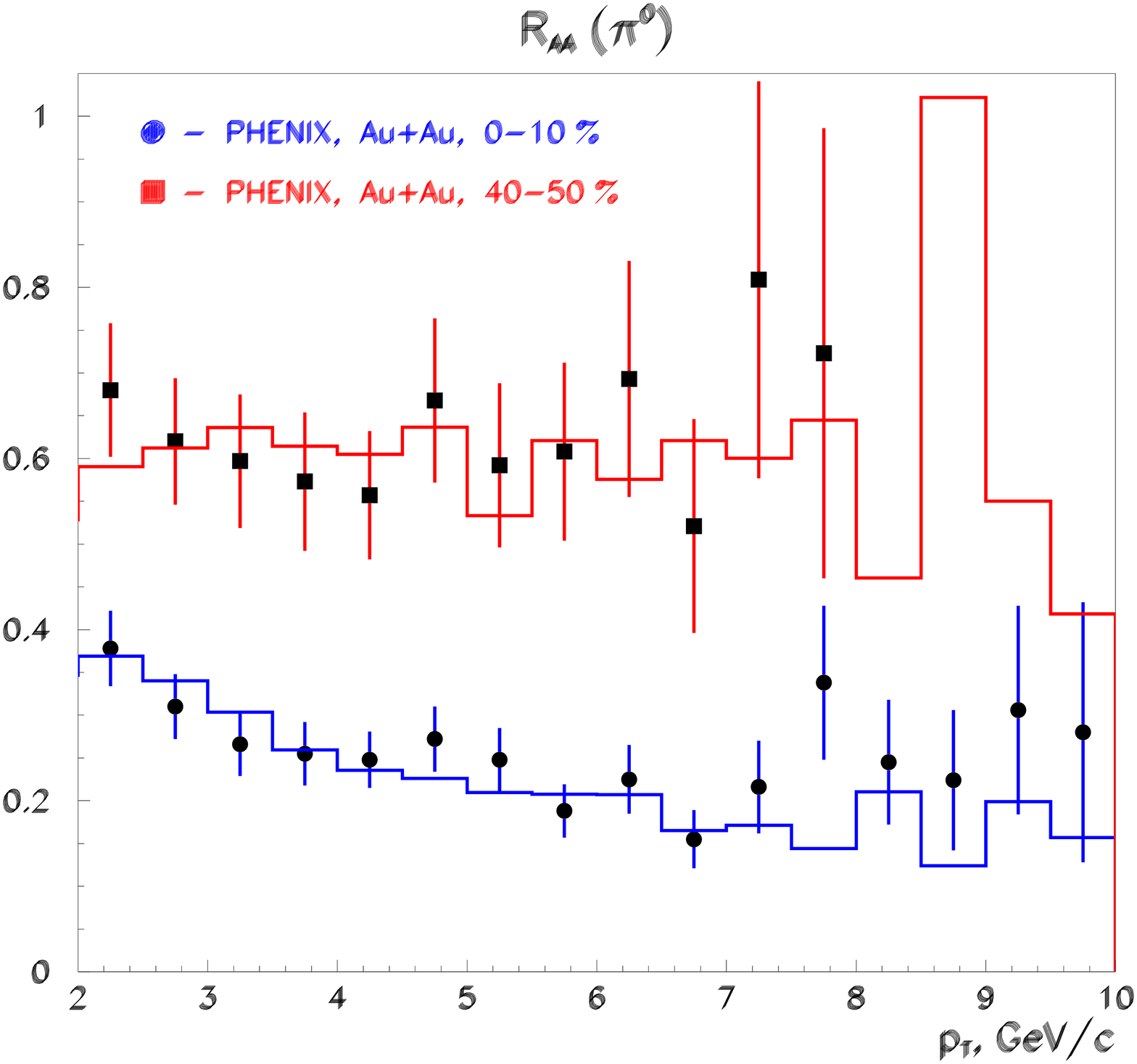,width=2.2in}}
\caption{The nuclear modification factor $R_{AA}$ for 
neutral pions in Au+Au collisions for two centrality sets. The points are  
PHENIX data, histograms are the model calculations.}
\label{fig3}
\end{figure}

\begin{figure}[b]
\centerline{\psfig{file=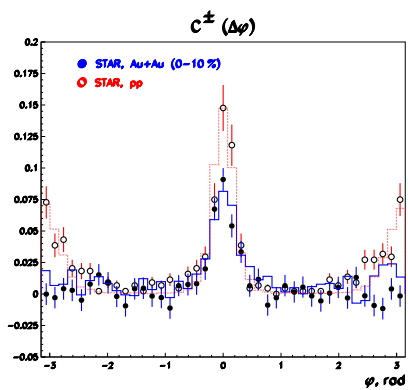,width=2.2in}}
\caption{The azimuthal two-particle correlation function for $pp$ and for 
central Au+Au collisions. The points are STAR data, dashed and 
solid histograms are the model calculations for $pp$ and Au+Au events 
respectively.}
\label{fig4}
\end{figure}

In order to demonstrate the efficiency of the model, the jet quenching pattern 
in Au+Au collisions at RHIC was considered. The PHOBOS data on $\eta$-spectra 
of charged hadrons~\cite{phobos} have been analyzed at first to fix the particle 
density in the mid-rapidity and the maximum longitudinal flow rapidity, 
$Y_L^{\rm{max}}=3.5$ (figure 1). To calculate multi-jet production cross 
section, the higher order corrections factor $K=2$ was used. The rest of the 
model parameters have been obtained by fitting PHENIX data on $p_T$-spectra of 
neutral pions~\cite{phenix} (figure 2): the kinetic freeze-out temperature 
$T_f=100$ MeV, maximum transverse flow rapidity $Y_T^{\rm{max}}=1.25$ and 
$p_T^{\rm min}=2.8$ GeV/$c$. The nuclear modification of the hardest domain of 
$p_T$-spectrum was used to extract initial QGP conditions: $T_0=500$ MeV and 
$\tau_ 0=0.4$ fm/$c$. Figure 3 shows that our model reproduce $p_T$-- and 
centrality dependences of nuclear modification factor $R_{AA}$ (determined as a 
ratio of particle yields in $AA$ and $pp$ collisions normalized on the number of 
binary NN sub-collisions) quite well.

Another important tool to verify jet quenching is two-particle azimuthal 
correlation function $C(\Delta \varphi)$ -- the distribution over an azimuthal 
angle of high-$p_T$ hadrons in the event with $2$ GeV/$c<p_T<p_T^{\rm trig}$  
relative to that for the hardest ``trigger'' particle with $p_T^{\rm trig}>4$ 
GeV/$c$. Figure 4 presents $C(\Delta \varphi)$ in $pp$ and in central Au+Au 
collisions (data from STAR~\cite{star}). Clear peaks in $pp$ collisions at 
$\Delta \varphi = 0$ and $\Delta \varphi =\pi$ indicate a typical dijet event 
topology. However, for central Au+Au collisions the peak near $\pi$ disappears. 
It can be interpreted as the observation of monojet events due to the 
absorption of one of the jets in a dense medium. Figure 4 demonstrates that 
measured suppression of azimuthal back-to-back correlations is well reproduced 
by our model.  

We leave beyond the scope of this paper the analysis of such important RHIC 
observables as the azimuthal anisotropy and particle ratios, which are 
sensitive to the soft physics. In order to study them, a more careful treatment 
of low-$p_T$ particle production than our simple approach is needed (the 
detailed description of space-time structure of freeze-out region, resonance 
decays, etc.). 

\section{Jet quenching at LHC}

The developed model was applied to analyze various novel features of jet 
quenching at the LHC (see~\cite{lokhtin-jff,lokhtin-bjet,lokhtin-mujet} for 
details). Let us just to enumerate some main issues. 

{\em Jets tagged by leading hadrons.} The relation between in-medium softening 
jet fragmentation function (measured in leading hadron channel) and suppression 
of jet rates due to energy loss out of jet cone, was analyzed in our 
work~\cite{lokhtin-jff}. We have found that the specific anti-correlation 
between two effects allows one to probe parton energy loss mechanism
(small-angular radiation vs. wide-angular radiation and collisional loss). 

{\em Jets induced by heavy quarks.} The possibility to observe the 
medium-modified fragmentation of hard b-quarks tagged by a energetic muon in 
heavy ion collisions was analyzed in the work~\cite{lokhtin-bjet}. We have found
that the reasonable statistics, $\sim 10^4$ events per one month of LHC run 
with lead beams, can be expected for the realistic geometrical acceptance and 
kinematic cuts. The significant softening b-jet fragmentation function 
determined by the absolute value of partonic energy loss and the angular 
radiation spectrum is predicted. 

{\em Z/$\gamma^*$+jet production.}  The channel with dimuon tagged jet 
production in heavy ion collisions was analyzed in the 
work~\cite{lokhtin-mujet}. The correlations between $\mu ^+\mu ^-$ pair and jet, 
as well as between $\mu ^+\mu ^-$ pair and a leading particle in a jet, were 
first numerically studied. We have found that the medium-induced partonic 
energy loss can result in significant $P_T$-imbalance between $\mu ^+\mu ^-$ 
pair and a leading particle in a jet, which is quite visible even for moderate 
loss. 

\section{Conclusions} 

The method to simulate jet quenching in heavy ion collisions has been 
developed. The model is the fast Monte-Carlo tool implemented to modify 
a standard PYTHIA jet event. The full heavy ion event is obtained as a 
superposition of a soft hydro-type state and hard multi-jets. The model is 
capable of reproducing main features of the jet quenching pattern at RHIC 
(the $p_T$ dependence of the nuclear modification factor and the suppression of 
azimuthal back-to-back correlations). The model was also applied to probe jet 
quenching in various new channels at LHC energy: jets tagged by leading 
particles, b-jets, dilepton-jet correlations. The further development of the 
model focusing on a more detailed description of low-$p_T$ particle production 
is planned for the future. 

\section*{Acknowledgments}
Discussions with A.~Morsch, C.~Roland, L.I.~Sarycheva, J.~Schukraft and 
B.~Wyslouch are gratefully acknowledged. This work is supported by grant 
N 04-02-16333 of Russian Foundation for Basic Research and Contract N
02.434.11.7074 of Russian Ministry of Science and Education.

\end{document}